\documentstyle[preprint,aps]{revtex}

\def\be{\begin{equation}}
\def\ee{\end{equation}}
\def\Be{\begin{eqnarray}}
\def\Ee{\end{eqnarray}}
\def\ba{\begin{array}}
\def\ea{\end{array}}

\def\ie{ i.\/e.\ }
\def\ncr{\nonumber\\}
\def\ve{\mbox{\boldmath $e$}}
\def\vep{\mbox{\boldmath $\epsilon$}}

\def\vd{\mbox{\boldmath $d$}}
\def\vk{\mbox{\boldmath $k$}}
\def\vepk{\mbox{\boldmath $\epsilon$} \cdot\mbox{\boldmath $k$}}
\def\vepd{\mbox{\boldmath $\epsilon$} \cdot\mbox{\boldmath $d$}}
\def\vkr{\mbox{\boldmath $k$} \cdot\mbox{\boldmath $r$}}
\def\bc#1{\left|\langle m|#1|l \rangle\right|}
\def\F#1#2{{\textstyle \frac{#1}{#2}}}
\def\r{\right}
\def\l{\left}
\def\f{\frac}
\def\w0{\omega_0}
\begin{document}
\draft
\title
{Modification of the Transition Rate in the Hydrogen Atom \\
placed in Finite Space}

\author{ Il-Tong Cheon \footnote{ On leave of absence from Department of Physics,
Yonsei University, Seoul 120-749, Korea.}
\footnote{ E.mail : itcheon@phya.yonsei.ac.kr} }

\address{Institute for Nuclear Study,
University of Tokyo, Tanashi 188 Tokyo, Japan}

\maketitle

\begin{abstract}
    When a hydrogen atom trapped in finite space formed by two parallel
perfectly conducting plates, the transition rates can be modified. Life-time of
the hydrogen $2P_{1/2}$-state is estimated to be shorter as much as $56.3ps $ with a
separation distance of $b = 1.2 \mu m$ between two plates. Although the modification
is dependent on position of the atom, the life-time can be shorten or lengthen by
adjusting the separation distance even for the case that the atom is placed at the
center between two plates.
\end{abstract}


\section*{ {\S}1. Introduction }

    A long time ago, Casimir[1] found that the zero-point energy appearing
in quantization of the classical radiation field in free space could be
observable by introducing two parallel perfectly conducting plates. He derived
\begin{equation}
\sum_{l j n} {1 \over 2} \hbar \omega_{ljn} -
  {b L^2 \over (2 \pi c)^3} \int_0^\infty {1 \over 2} \hbar \omega d^3 \omega
  = - {\pi^2 L^2 \over 720 b^3} \hbar c ,
\end{equation}
where $L^2$ is the area of plates and b is the separation distance of two
plates. The first and second terms on the left hand side denote the zero-point
energies of radiation fields quantized in finite and free spaces,
respectively. Notice the
non-vanishing value on the right-hand side which is inversely proportional to
cube of $b$. This result implies spatial dependence of the zero-point energy.
On the other hand, it was also shown that the atomic energy levels could be
modified when the atom was placed between two parallel plates [2,3,4].
Generally, the energy levels are expressed by real parts of the energy
eigen-values  while the life-times of energy levels, equivalently transition
rates, are given by imaginary values of the complex energy eigen-values.
Therefore, one may expect that the atomic transition rates will also be
modified as long as the energy levels are shifted in finite space[5,6].
In this
paper, we investigate the atomic transition rate in the framework similar to
Casimir's calculation of the attractive force exerting between two plates [1].

\section*{ {\S}2. Calculation of the transition rate}
\subsection*{2.1 Case of free space }

    Let us start to briefly review the calculation of transition rate of the
hydrogen atom in free space.

    The time-dependent perturbation theory in quantum mechanics gives the
first-order amplitude of spontaneous radiation as [7]
\begin{equation}
a_m^{(1)}(t) = {1 \over i \hbar} \int_0^t dt' {\langle m|} H_I(t') {| l \rangle}
    \exp \{ i(E_m - E_l) t' / \hbar \} ,
\end{equation}
where the interaction Hamiltonian is expressed as
\begin{equation}
H_I(t') = H_I^{'} \exp(\pm i \omega t').
\end{equation}
Then, the transition probability is
\begin{equation}
|a_m^{(1)}(t)|^2 = {1 \over \hbar^2} | {\langle m |} H_I^{'} {| l \rangle} |^2
                | I_{ml}^{(\pm)} |^2 ,
\end{equation}
where
\begin{equation}
I_{ml}^{(\pm)} = \int_0^t dt' \exp [i(\omega_0 \pm \omega) t']
\end{equation}
with $\hbar \omega_0 = E_m - E_l$. The negative(positive) sign in eq.(5)
denotes the photon emission(absorption). Hereafter, we consider only the photon
emission, {i.e.\/ } negative sign in eq.(5). Carrying out the integration directly,
we obtain
\begin{eqnarray}
|I_{ml}^{(-)}|^2 &=& | { \exp\{i(\omega_0 - \omega) t\} - 1
                         \over i (\omega_0 - \omega)        } |^2 \nonumber \\
                 &=& 4 { \sin^2[ {t \over 2} (\omega_0 - \omega)]
                          \over (\omega_o - \omega)^2 }.
\end{eqnarray}
Then, the transition rate is given by [7,8]
\begin{eqnarray}
W_0 &=& \frac{\partial}{\partial t} \int  |a_m^{(1)}(t)|^2 \rho d(\hbar \omega) \nonumber \\
    &=& {2 \over \hbar^2} \int | {\langle m |} H_I^{'} {| l \rangle} |^2
        { {\sin[t(\omega_0 - \omega)]} \over {(\omega_0 - \omega)}}
        { {V \omega^2} \over {(2\pi)^3 c^3} }
        d\Omega d\omega.
\end{eqnarray}
If the dipole approximation is taken for $H^{'}_I$, eq.(7) can be given as
\begin{eqnarray}
W_0 &=& {2 \over \hbar^2} {\hbar \over 2 V}
        \int \sum_i | {\langle m |} e^{-i \mbox{\boldmath $k$} \cdot \mbox{\boldmath $r$}_i}
                      \mbox{\boldmath $\epsilon$} \cdot \mbox{\boldmath $d$}_i {| l \rangle} |^2
             { \sin[t (\omega_0 - \omega)]
               \over (\omega_0 - \omega)}
             { V \omega^3 \over (2 \pi c)^3 }
        d\Omega d\omega \nonumber \\
    &=& { e^2 \over 8 \pi^3 \hbar c^3 }
        \int | {\langle m |} \mbox{\boldmath $r$} {| l \rangle} \cdot \mbox{\boldmath $\epsilon$} |^2 d\Omega
        \int_{-\infty}^\infty { \omega^3 \sin[t (\omega_0 - \omega)]
                                \over (\omega_0 - \omega)} d\omega,
        \eqnum{7a}
\end{eqnarray}
where $\mbox{\boldmath $\epsilon$}$ is the polarization vector and
$\mbox{\boldmath $d$} = e \mbox{\boldmath $r$}$ is a dipole moment. Considering that
\begin{equation}
{\langle m |} \mbox{\boldmath $r$} {| l \rangle} \cdot \mbox{\boldmath $\epsilon$} =
\mbox{\boldmath $r$}_{ml}\cos \Theta,
\end{equation}
and [7]
\begin{equation}
\int \cos^2 \Theta d\Omega = {8 \pi \over 3},
\end{equation}
we find
\begin{eqnarray}
W_0 &=& {e^2  \over 3 \pi^2 \hbar c^3} |\mbox{\boldmath $r$}_{ml}|^2
        \int_{-\infty}^\infty { \omega^3 \sin[t(\omega_0 - \omega)]
                                \over (\omega_0 - \omega)}
        d\omega
      \nonumber \\
    &=& \left( {e^2 \over 4 \pi \hbar c} \right) {4 \over 3}
        {\omega_0^3 \over c^2} |\mbox{\boldmath $r$}_{ml}|^2 .
\end{eqnarray}
The matrix element $\mbox{\boldmath $r$}_{ml}$ can be calculated with the hydrogen wave
functions and the result for the $2P_{1/2} \rightarrow 1S_{1/2}$ transition is
known as
\begin{equation}
|\mbox{\boldmath $r$}_{2P,1S}|^2 = {1\over3}
    \left( a_0 {16 \over \sqrt{6}} \left({2\over3}\right)^4 \right)^2 ,
\end{equation}
where $a_0$ is the Bohr radius. Since $\nu(2S_{1/2} \rightarrow 1S_{1/2}) =
2466061395.6(4.8) MHz$[9] and the Lamb shift yields
$\nu(2S_{1/2} \rightarrow 2P_{1/2}) = 1057.8594 \pm 0.0019 MHz$ [10],
we find $\nu(2P_{1/2} \rightarrow 1S_{1/2}) = \omega_0(2P_{1/2} \rightarrow 1S_{1/2})/2\pi =
2466060337.7  MHz$ and, then, the transition rate is
\begin{equation}
W_0 (2P_{1/2} \rightarrow 1S_{1/2}) = 0.626288 \times 10^9 Hz,
\end{equation}
which gives the life-time of the $2P_{1/2}$-state
\begin{equation}
\tau_0(2P_{1/2}) = \left[ W_0(2P_{1/2} \rightarrow 1S_{1/2}) \right]^{-1}
    = 1.597 \times 10^{-9} s.
\end{equation}
This is the result for the hydrogen atom in free space.

\subsection*{2.2 Coordinate system of dipole moment and polarization vector}

Although the polarization vector, $\vep$, is perpendicular to the wave vector, $\vk$,
of the radiation field, \ie $\vepk =0$, the dipole moment,$\vd$, is generally not
specified in any direction. It is better to clarify relations among these vectors.

First of all, the space-fixed coordinates are assigned by unit vectors ($\ve_1 ,\ve_2 ,\ \ve_3 $)
and the wave vector of radiation field, $\vk$, is chosen in the direction of a unit
vector $\bar{\ve}_3$. Here, unit vectors ($\bar{\ve}_1 ,\ \bar{\ve}_2 ,\ \bar{\ve}_3 $)
are orthogonal to each other. Then, we have
\begin{equation}
\bar{\ve}_3 =\ve_1 \sin \theta \cos \phi + \ve_2 \sin \theta \sin \phi +\ve_3 \cos \theta
\eqnum{15a}
\label{15a}
\end{equation}
where $\theta$ is the angle vetween $\ve_3$ and $\bar{\ve}_3$ and $\phi$ is the angle
between $\ve_1$ and a projection of $\bar{\ve}_3$ on the $\ve_1 -\ve_2$ plane. Since
$\bar{\ve}_1$ is perpendicular to $\bar{\ve}_3$, replacing $\theta$ by
$\theta + \textstyle{\pi \over 2}$ in eq.(\ref{15a}), we find
\begin{equation}
\bar{\ve}_1 =\ve_1 \cos \theta \cos \phi + \ve_2 \cos \theta \sin \phi -\ve_3 \sin \theta.
\eqnum{15b}
\label{15b}
\end{equation}
Considering that $\bar{\ve}_2$ is perpendicular to the $\bar{\ve}_1 - \bar{\ve}_3$
plane, we replace $\phi$ by $\phi +\textstyle{\pi \over 2}$ and set
$\theta=\textstyle{\pi \over 2}$ in eq.(\ref{15a}). Then, we can have
\begin{equation}
\bar{\ve}_2=-\ve_1 \sin\phi + \ve_2 \cos\phi.
\eqnum{15c}
\label{15c}
\end{equation}
Polarization of the radiation field is perpendicular to $\vk$, \ie, $\vepk =0$
and, then, it is expressed as
\setcounter{equation}{15}
\begin{equation}
\vep_i =\bar{\ve}_i \ \ \ (i=1,\ 2).
\label{16}
\end{equation}
If the dipole moment is along $\ve_3$, \ie $\vd=d \ve_3$, we find by
eqs.(\ref{15b}) and (\ref{15c})
\be
\int_0^{2 \pi} |< m|(\vepd)|l>|^2 d\phi =  \int_0^{2 \pi}d_{ml}^2\sin^2\theta d\phi
= 2\pi d_{ml}^2(1-\cos^2\theta ) =2 \pi d_{ml}^2\left(1-\frac{k_z^2}{k^2}\right).
\ee
This result yields that given in eq.(4.9) of ref.11. On the other hand, for
$\vd =d \ve_1$,
\Be
\int_0^{2 \pi} |< m|(\vepd)|l>|^2 d\phi &=& d_{ml}^2 \int_0^{2 \pi}[(\cos\theta \cos\phi)^2
+(\sin\phi)^2 ] d\phi \ncr
&=& \pi d_{ml}^2 (1+\cos^2\theta ) =\pi d_{ml}^2 \left(1+\frac{k_z^2}{k^2}\right),
\Ee
which yields the result given in eq.(4.17) of ref.11. For $\vd= d \ve_2$, we
obtain the same result as eq.(18),
\be
\int_0^{2 \pi} |<m|(\vepd)|l>|^2 d\phi =\pi d_{ml}^2 (1+\cos^2\theta ) =
\pi d_{ml}^2 \left( 1+\frac{k_z^2}{k^2}\right) .
\ee

Let us explore a general case that $\vd$ is not polarized along a specific direction,
\ie $\vd (\theta', \phi')$. When the normal mode of the electric field is expressed
$u_j \ (j=x,\ y,\ z)$, we have
\Be
|<m|(\vepd) u|l>|^2 &\equiv& \sum_i |<m| \epsilon_{ix} d_x u_x +\epsilon_{iy}d_y d_y +
\epsilon_{iz}d_z u_z |l>|^2 \ncr
&=&d_{ml}^2 \{|u_x \cos\theta\cos\phi\sin\theta'\cos\phi' +u_y \cos\theta\sin\phi\sin\theta'\sin\phi'-
u_z \sin\theta\cos\theta'|^2 \ncr
&&\ + |-u_x \sin\phi\sin\theta'\cos\phi' + u_y \cos\phi\sin\theta'\sin\phi'|^2 \}.
\Ee
Integration over $\phi$ yields
\Be
\int_0^{2 \pi} |<m|(\vepd)u|l>|^2 d\phi &=&\pi d_{ml}^2 \{|u_x|^2 (1+\cos^2\theta)\sin^2\theta'
\cos^2\phi'\ncr
&& \ + |u_y|^2 (1+\cos^2\theta )\sin^2 \theta' \sin^2\phi' +2 |u_z|^2\sin^2\theta
\cos^2\theta' \},
\Ee
which becomes
\be
\int_0^{2 \pi} |<m|(\vepd)u|l>|^2 d\phi = \pi d_{ml}^2 \{ |u_{xy}|^2 (1+\cos^2\theta )
\sin^2\theta' + 2 |u_z|^2 \sin^2\theta \cos^2\theta' \}
\ee
under an assumption of $u_x =u_y \equiv u_{xy}$. Notice that eq.(22)
does not depend on $\phi'$ any more. When $u_{xy}=u_z=1$ and $\theta'=0$, \ie $\vd$
is along the z direction, eq.(22) leads to eq.(17), while with $u_{xy}=u_z=1$ and
$\theta'={\textstyle \frac{\pi}{2}}$, \ie $\vd$ is parallel to the $x$- or $y$-axis,
eq.(22) leads to eq.(18). Furthermore, integration of eq.(22) over $\theta$ gives
\be
\int_0^\pi \int_0^{2 \pi} |<m| (\vepd) u|l>|^2 \sin\theta d\theta d\phi =
\pi d_{ml}^2 \left\{\frac{8}{3} \left|u_{xy}\right|^2\sin^2\theta' +\frac{8}{3}
\left|u_z\right|^2 \cos^2\theta'\right\},
\ee
which yields $\textstyle{{8\over 3}} \pi d_{ml}^2$, when $u_{xy}=u_z =1$. Actually, eqs.(9) and
(10) were obtained by this manner.

\subsection*{ 2.3 Case of finite space}

In  finite space, the electromagnetic field has to be quantized to satisfy the
boundary condition at the wall. And the energy-momentum tensor is constructed with
this quantized field which is a solution of the Klein-Fok equation[12].

Since the electromagnetic field has a distinctive feature of transverse, the
boundary condition is imposed at the wall such that the tangential components of the
electric field vanish there. The simplest modification of photon propagator
is to discretize the electromagnetic field modes in the direction perpendicular
to the wall[13]. This photon propagator can be proved[14] to be an approximation
of the more reliable photon propagator derived by direct insertion of the
$\delta$~-~function to satisfy the boundary condition[15].

All these discussions lead to a result that the energy-momentum tensor gives the
discretized energy in the vacuum state as
\begin{eqnarray}
E=\int_0^b d\!x<0|T_{00}|0>= \frac{1}{2}\sum_n \omega_n
\end{eqnarray}
Therefore, the photon emitted spontaneously in finite space must have a
discrete energy instead of the continuous one. If the hydrogen is
placed between two parallel plates of size $L \times L $ seperated from each
other by $b(b<<L)$, $z$-component of the momentum of the radiation field which is
perpendicular to the plate should be discretized and the integral over it has
to be replaced by a summation. The result obtained by such a replacement is exactly
identical with that derived by direct quantization of the electromagnetic field in finite
space[11,16,17]. Compare eq.(7a) to eq.(4.7) of ref.11. Therefore, without
repeating the field quantization in finite space, we shall start from eq.(7a) to obtain
the transition rate in finite space. Nevertheless, our theory is rigorously based
on the quantization given in ref.11.

In addition to such a replacement, we must change $k_z$ to ${\textstyle \frac{\pi}{b} n}$
$(n=0,\ \pm1,\ \pm2,\ \cdots)$ and [11]
\be
\frac{1}{\sqrt{V}}\exp(-i\vkr)\rightarrow
\left\{ \ba{rcl}
 \frac{1}{\sqrt{V}}u_z &=&\sqrt{\frac{2}{V}}\exp[-i (k_x x+k_y y)]\cos(k_z z)\\
       \frac{1}{\sqrt{V}}u_{x,y} &=&\sqrt{\frac{2}{V}}\exp[-i (k_x x+k_y y)]i\sin(k_z z)
    \ea\right.
\ee
where a factor $\sqrt{2}$ comes from normalization.

Thus, the transition rate in finite space can be obtained from
\be
W=\frac{2}{\hbar^2}\frac{\hbar}{2 V} \int d\Omega\int_{-\infty}^\infty d \omega |<m|(\vepd)\bar{u}(k_z R_z)|l>|^2
\frac{\omega^3 \sin[t(\omega_0-\omega)]}{(\omega_0-\omega)}\frac{V}{(2 \pi c)^3},
\ee
where $\bar{u}_x (k_z R_z)=\bar{u}_y(k_z R_z)=i\sin(k_z R_z ) $ and $\bar{u}_z (k_z R_z)=\cos(k_z R_z)$.
$R_z (0<R_z \le {\textstyle \frac{b}{2}})$ is a distance between the atom and one of
these plates. It should be reminded here that $\omega =ck=c (k_x^2+k_y^2 +k_z^2)^{1/2}$
and the integral over $k_z$ is replaced by a summation over $n$ because of
$k_z ={\textstyle \frac{\pi}{b} n}$. Using the relation
\be
\int\!\!\!\int\limits^\infty_{ -\infty}\!\!\!\int d k_x dk_y dk_z \frac{1}{k}=
\int d\phi\int d\theta \sin\theta \int_0^\infty dk k =
\frac{1}{c^2} \int d\phi\int d\theta \sin\theta\int_0^\infty d\omega \omega
\ee
and replacing the integral over $k_z$ by a summation, we find the integral in eq.(26)
as
\Be
I&\equiv& \int d\Omega\int_{-\infty}^\infty d \omega |<m|(\vepd)\bar{u}(k_z R_z)|l>|^2
\frac{\omega^3 \sin [t(\omega_0-\omega)]}{(\omega_0-\omega)} \ncr
&=&\int d\Omega\int_0^\infty d \omega |<m|(\vepd)\bar{u}(k_z R_z)|l>|^2
\left\{\frac{\omega^3 \sin[t(\omega_0-\omega)]}{(\omega_0-\omega)}
-\frac{\omega^3 \sin[t(\omega_0+\omega)]}{(\omega_0+\omega)}\right\}, \ncr
&\Rightarrow&\frac{\pi c^4}{b} \int_{-\infty}^\infty\int_{-\infty}^\infty dk_x dk_y \sum_{n=-\infty}^{\infty}
\bc{(\vepd)\bar{u}(\frac{\pi n}{b} R_z )}^2  \ncr
&& \ \times \left\{ \frac{k\sin[t (\omega_0 -ck)]}{(\omega_0-ck)}
-\frac{k\sin[t (\omega_0 +ck)]}{(\omega_0+ck)}\right\},
\Ee
where $k=[k_x^2+k_y^2+(\frac{\pi}{b} n)^2 ]^{1/2}$. By transforming variables,
$k_x$ and $k_y$, into $(\F{\pi}{L} i)$ and $(\F{\pi}{L} j)$, equation (28) becomes
\Be
I&=&c^3 \left(\frac{\pi}{L}\right)^2 \left(\frac{\pi}{b}\right) \int_{-\infty}^\infty
\int_{-\infty}^\infty di dj \sum_{n=-\infty}^\infty \bc{(\vepd)\bar{u}\left(\frac{\pi R_z}{b} n\right)}^2\ncr
&&\ \times\left\{\frac{B \sin[t\omega_0 (1-B)]}{(1-B)}-\frac{B\sin[t\omega_0 (1+B)]}{(1+B)}\right\},
\Ee
where
\be
B=\l[ \l(\frac{c\pi}{\omega_0 L} i\r)^2 +\l(\frac{c \pi}{\omega_0 L} j\r)^2
+\l(\frac{c \pi}{\omega_0 b} n\r)^2\r]^{1/2}.
\ee
Again changing variables as $i=(\F{\omega_0 L}{c \pi}) x$ and $j=(\F{\omega_0 L}{c\pi}) y$, we
arrive at
\Be
I&=&\frac{c \pi\omega_0^2}{b}\int_{-\infty}^\infty \int_{-\infty}^\infty dx dy
\sum_{n=-\infty}^\infty \bc{(\vepd)\bar{u}(\frac{\pi R_z}{b} n)}^2 \ncr
&&\ \times\l\{\frac{B_n \sin[t\omega_0 (1-B_n)]}{(1-B_n)}-\frac{B_n \sin[t\omega_0 (1+B_n)]}{(1+B_n)}\r\},
\Ee
where
\be
B_n =\l[x^2+y^2+\l(\frac{c\pi}{\omega_0 b}n\r)^2\r]^{1/2}.
\ee
In order to evaluate the summation over n, it is convenient to make use of
Poisson's summation formula on Fourier transforms,
\be
\sum_{n=-\infty}^\infty f(n)=\sum_{s=-\infty}^\infty \int_{-\infty}^\infty f(u) \exp(2\pi i su) du.
\ee
Then, eq.(31) can be expressed as
\Be
I^{(\pm)}&=&\frac{c\pi\omega_0^2}{b}\sum_{s=-\infty}^\infty
\int\!\!\!\int\limits^\infty_{-\infty}\!\!\!\int dx dy du \bc{(\vepd) \bar{u}\l(\frac{\pi R_z}{b} u\r)}^2\ncr
&&\ \times\frac{\sqrt{x^2+y^2+\l(\frac{c\pi}{\omega_0 b} u\r)^2}\sin\l\{t\omega_0
\l(1\pm\sqrt{x^2+y^2+\l(\frac{c\pi}{\omega_0 b} u\r)^2}\r)\r\}}{\l\{1\pm\sqrt{x^2+y^2+
\l(\frac{c\pi}{\omega_0 b} u\r)^2}\r\}}
\exp (2\pi i su)\ncr
&=&\omega_0^3 \sum_{s=-\infty}^\infty
\int\!\!\!\int\limits^\infty_{-\infty}\!\!\!\int dx dy dz \bc{(\vepd) \bar{u}\l(\frac{\omega_0 R_z}{c} z\r)}^2\ncr
&&\ \times\frac{\sqrt{x^2+y^2+z^2}\sin\l\{t\omega_0
\l(1\pm\sqrt{x^2+y^2+z^2}\r)\r\}}{\l(1\pm\sqrt{x^2+y^2+z^2}\r)}
\exp \l(i\frac{2\omega_0 bs}{c} z\r)\ncr
&=&\omega_0^3 \sum_{s=-\infty}^\infty
\int d\Omega \int_0^\infty dr \bc{(\vepd) \bar{u}\l(\frac{\omega_0 R_z}{c} r \cos\theta \r)}^2\ncr
&&\ \times\frac{r^3\sin\l\{t\omega_0
\l(1\pm r \r) \r\} }{\l(1\pm r \r)}
\exp \l(i\frac{2\omega_0 bs}{c} r \cos\theta \r).
\Ee
Therefore, we obtain
\Be
I&=&\omega_0^3 \sum_{s=-\infty}^\infty
\int  d\Omega \int_0^\infty dr \bc{(\vepd) \bar{u}\l(\frac{\omega_0 R_z}{c} r \cos\theta \r)}^2\ncr
&&\ \times\l\{ \frac{r^3\sin\l\{t\omega_0\l(1 - r \r) \r\} }{\l(1- r \r)}
-\frac{r^3\sin\l\{t\omega_0\l(1 + r \r) \r\} }{\l(1+ r \r)} \r\}
\exp \l(i\frac{2\omega_0 bs}{c} r \cos\theta \r)\ncr
&=&\omega_0^3 \sum_{s=-\infty}^\infty
\int  d\Omega \int_{-\infty}^\infty dr \bc{(\vepd) \bar{u}\l(\frac{\omega_0 R_z}{c} r \cos\theta \r)}^2\ncr
&&\ \times \frac{r^3\sin\l\{t\omega_0\l(1 - r \r) \r\} }{\l(1- r \r)}
\exp \l(i\frac{2\omega_0 bs}{c} r \cos\theta \r)\ncr
&=&\pi \omega_0^3 \sum_{s=-\infty}^\infty
\int  d\Omega \bc{(\vepd) \bar{u}\l(\frac{\omega_0 R_z}{c} \cos\theta \r)}^2
\exp \l(i\frac{2\omega_0 bs}{c} \cos\theta \r),
\Ee
where we used
\be
\frac{\sin[t\omega_0 (1-r)]}{(1-r)}\simeq \pi \delta (1-r)
\ee
for $\omega_0 t$ being very large, \ie $\omega_0 =1.5\times 10^{16} s^{-1}$ and
$t\simeq 1.6\times 10^{-9}s$ for the hydrogen $2 P_{1/2} \rightarrow 1
S_{1/2}$ transition. By
eq.(22), I can be given as
\Be
I&=&\pi^2 \omega_0^3 d_{ml}^2 \int_0^\pi d\theta \sin\theta
\l\{ 2 \sin^2 \l( \frac{\omega_0 R_z}{c} \cos\theta
                \r)  \l( 1+\cos^2\theta
                     \r) \sin^2\theta'
 \r. \ncr
& & \ \l. +4 \cos^2 \l( \frac{\omega_0 R_z}{c}\cos\theta
                    \r)  \l( 1-\cos^2\theta
                         \r) \cos^2\theta'
      \r\} \sum_{s=-\infty}^\infty \exp \l[ i \frac{2 \omega_0 bs}{c}\cos\theta
                                        \r].
\Ee
By integrating over $\theta$, eq.(37) is found in the form
\Be
I&=&\pi^2 \omega_0^3 d_{ml}^2 \sum_{s=-\infty}^\infty
 {\Big\{}\sin^2\theta' [4 j_0 (\xi_2 )-2 j_0 (\xi_1 +\xi_2 )-2 j_0 (\xi_1 -\xi_2 )\ncr
 &-& \frac{4}{\xi_2} j_1 (\xi_2 ) +\frac{2}{(\xi_1 +\xi_2 )} j_1 (\xi_1 +\xi_2 ) +
     \frac{2}{(\xi_1 -\xi_2 )} j_1 (\xi_1 -\xi_2 )] \ncr
 &+& \cos^2 \theta'[\frac{8}{\xi_2} j_1(\xi_2) + \frac{4}{(\xi_1+\xi_2)}
     j_1(\xi_1+\xi_2)+\frac{4}{(\xi_1-\xi_2)} j_1(\xi_1-\xi_2)]{\Big\}},
\Ee
where $j(\xi)$ is the spherical Bessel function, $\xi_1 =\F{2 \omega_0 R_z}{c}$ and
$\xi_2 =\frac{2 \omega_0 b}{c} s$. Thus, the transition rate is given as
\Be
W&=&\frac{1}{8 \pi^3 \hbar c^3} I
 = W_1 +W_2 (b),
 \Ee
 where
 \Be
& &W_1=W_0 \l\{ \sin^2 \theta' \l[ 1-\frac{3}{2} j_0 (\xi_1)+\frac{3}{2 \xi_1} j_1 (\xi_1 )\r]
     +\cos^2 \theta' \l[ 1+\frac{3}{\xi_1} j_1 (\xi_1 )\r]\r\},\\
& &W_2(b) = W_0 \l\{ \sin^2 \theta' \sum_{s=1}^\infty \l[ 3 j_0 (\xi_2 )
                      -\frac{3}{2} j_0 (\xi_1 +\xi_2 ) -\frac{3}{2} j_0 (\xi_1 -\xi_2 ) \r. \r. \ncr
         & & \ \ \ \ \ \ \ \  \l.-\frac{3}{\xi_2} j_1 (\xi_2 ) +\frac{3}{2 (\xi_1 +\xi_2 )} j_1 (\xi_1 +\xi_2 )
           + \frac{3}{2(\xi_1 -\xi_2 )} j_1 (\xi_1 -\xi_2 )\r] \ncr
         & &\ \ \ \l. + \cos^2 \theta' \sum_{s=1}^\infty \l[ \frac{6}{\xi_2} j_1 (\xi_2 )+
         \frac{3}{(\xi_1 +\xi_2 )} j_1 (\xi_1 +\xi_2 )
         +\frac{3}{(\xi_1 -\xi_2 )} j_1 (\xi_1 -\xi_2 )\r]\r\} \ .
\Ee
$W_1$ denotes the $s=0$ term, and it is independent of $b$, while $b$ dependence
appears only in $W_2 (b)$. Particularly, when $\vd$ is polarized along the $x$ or $y$
direction, \ie $\theta' =\F{\pi}{2}$, $W_1 $ leads to
\be
W_{1x} =W_{1y} =W_0 \l[ 1-\frac{3}{2} j_0 (\xi_1 ) +\frac{3}{2 \xi_1} j_1 (\xi_1 )\r].
\ee
If $\vd$ is along the $z$ direction, \ie $\theta' =0$, it becomes
\be
W_{1z}=W_0 \l[ 1+\frac{3}{\xi_1}j_1 (\xi_1 )\r].
\ee
These results are exactly the same as those given in refs. 11 and 16.

By defining
\Be
P_1 &=& {{3 j_1(\frac{2 \omega_0 b}{c}({1 \over 2} - {Z \over b}))}
   \over{ (\frac{2 \omega_0 b}{c}({1 \over 2} - {Z \over b}))}},\\
P_2 &=& - {3 \over 2} j_0(\frac{2 \omega_0 b}{c}({1 \over 2} - {Z \over b}))
  + {1 \over 2} P_1,\\
Q_1 (b)&=&\sum_{s=1}^\infty \l[ 3 j_0 \l(\frac{2 \omega_0 b}{c} s \r) -
\frac{3}{2}j_0 \l(\frac{2\omega_0 b}{c} \l(s+\frac{1}{2} -\frac{Z}{b} \r)\r)-
\frac{3}{2} j_0 \l( \frac{2 \omega_0 b}{c} \l( s-\frac{1}{2} +\frac{Z}{b} \r)\r) \r.\ncr
& & \ \ \  \l.  - \frac{3 j_1 \l( \f{2 \w0 b}{c} s \r)}{ \l( \f{2 \w0 b}{c} s \r)} +
\f{3 j_1 \l( \f{2 \w0 b}{c} \l(s+\f{1}{2}-\f{Z}{b}\r)\r)}{2 \l( \f{2 \w0 b}{c}\r)
\l(s+\f{1}{2}-\f{Z}{b}\r)}+
\f{3 j_1 \l( \f{2 \w0 b}{c} \l(s-\f{1}{2}+\f{Z}{b}\r)\r)}{2 \l( \f{2 \w0 b}{c}\r)
\l(s-\f{1}{2}+\f{Z}{b}\r)} \r],\\
Q_2 (b)&=&\sum_{s=1}^\infty \l[ \f{6 j_1 \l(\frac{2 \omega_0 b}{c} s \r)}{\l(\frac{2 \omega_0 b}{c} s \r)}
 +\f{3 j_1 \l( \f{2 \w0 b}{c} \l(s+\f{1}{2}-\f{Z}{b}\r)\r)}{ \l( \f{2 \w0 b}{c}\r)
\l(s+\f{1}{2}-\f{Z}{b}\r)}+
\f{3 j_1 \l( \f{2 \w0 b}{c} \l(s-\f{1}{2}+\f{Z}{b}\r)\r)}{ \l( \f{2 \w0 b}{c}\r)
\l(s-\f{1}{2}+\f{Z}{b}\r)}\r],
\Ee
the transition rate can be written as
\be
W=W_0 [ 1+\sin^2\theta' \lbrace P_2 + Q_1 (b) \rbrace +\cos^2 \theta' \lbrace
P_1 + Q_2 (b) \rbrace ].
\ee
Here $Z=\F{b}{2}-R_z \ (0\leq Z < \F{b}{2} )$ is a distance of the atom in
the $z$-direction from the center of finite space.

Since the dipole moment is not polarized in the specific direction in a practical
experiment, we should take average values, $<\sin^2 \theta' >=<\cos^2 \theta' >=1/2$, and,
then
\be
\bar{W}=W_0 \{1 + \Delta_0 + \Delta (b) \}
\ee
with
\Be
\Delta_0 = \f{1}{2}(P_1 + P_2),\\
\Delta (b) =\f{1}{2} [Q_1 (b) +Q_2 (b) ].
\Ee
For $ b \rightarrow \infty $, the last term in eq.(49) vanishes
but the second term, $\Delta_0$, can survive as far as $R_z$ is small,
i.e. the atom is located closely to one of the walls. Accordingly,
for the case of one wall, the modification can be given by
$\Delta_0 $ - term.

\section*{\S3. Numerical results}

With various values of $b$ and $Z$, we can obtain numerical values of the shift of
transition rate by eq.(49). Our results are shown in Table 1 for the hydrogen
$2 P_{1/2}\rightarrow 1 S_{1/2}$ transition with $Z=0$. It can be seen that sign of
the shift changes in accordance with $b$ value. We define
\be
\Delta W=\bar{W} -W_0 =W_0 \lbrace \Delta_0 + \Delta (b) \rbrace,
\ee
values of which are also listed in Table 1.

For $Z=0$ (notice $0\leq Z<\F{b}{2}$) and $b=1.2\mu m$, we obtain $\Delta_0 =
+0.00442, \Delta (b) =+0.0321 $ and $\Delta W =+22.9$MHz,
while $\Delta W=+38.2$MHz for $b=0.7 \mu m$.
They are in a measureable range because the accuracy of the present technique is 4.8MHz[9].

The life-time is given by inverse of the transition rate,
\be
\Delta \tau =\tau -\tau_0
\ee
with $\tau_0 =1.597 $ns. Our results are listed in Table 1 and 2. Particularly, Table 2
shows dependence of the transition rate on position of the atom, when $b=1.2\mu m$.
At $Z=b/10$ which is very close to the center, $\Delta_0 + \Delta (b)$ yields
+0.036, \ie
$\Delta W=+22.8$MHz. This is corresponding to $\Delta \tau =-56.1 ps$.

\section*{\S4. Conclusion}

In this paper, we have calculated effects of the finite space on the transition
rate and the life-time of the hydrogen atom.

We have derived explicitly $b$ dependence of the transition rate, $\Delta (b)$,
which was not evaluated in the previous calculations [11,17]. The result shows the
characteristic oscillation of the transition rate in accordance with $b$ values.
This property depends very much on value of the frequency $\omega_0$.  The
magnitude of $\Delta (b)$ varies with value of $\omega_0$.  $b$-dependence
is around $2\sim 5 \%$ for a separation distance of $\mu m$ order. However, it is
much enhanced for smaller value of $\w0$. For example, with $\w0 =1.2\times 10^{11}$Hz,
we find that contribution of the $b$-dependence term, $\Delta (b)$, is 33\% for $b=1 cm$.

As is seen in Table 2, variation of the transition rate and life-time are very sensitive
to position of the atom. Since it is quite hard to set up the atom at an accurate
position, measurement of the absolute magnitudes of $\Delta W$ and $\Delta\tau$ may be
very difficult. However, observation of these shifts seems to be definitely
possible.

\acknowledgements

This work was supported by the Korean Ministry of Education(Contract no.
BSRI-96-2425) and the Korean Science and Engineering Foundation.
I thank T. Yamazaki and Y. Akaishi
for their warm hospitality extended at Institute for Nuclear Study,
University of Tokyo.

\begin{table}
\noindent Table 1. Shifts of the transition rate and the life-time for $Z=0$.
 $\w0=1.5\times 10^{16}$Hz.
\begin{tabular}{ccccccc}
$b(\mu m)$& $Q_1 (b) \times 10^2 $ &$Q_2 (b) \times 10^3 $ &
$\Delta_0 \times 10^2$ &
$\Delta (b)\times10^2$ & $\Delta W$(MHz) & $\Delta \tau$(ps) \\ \hline
0.5 & -0.1532  & -9.127   & 0.038       & -0.5330 & -3.10 & 7.94 \\ \hline
0.6 & 2.053    & 0.9236   & 2.423         &1.073 & 21.90 & -53.95 \\ \hline
0.7 & 9.865    & 1.556    & 1.081         &5.010 & 38.15 & -91.69 \\ \hline
0.8 & -5.800   & 1.196    & -1.301      &-2.840 & -25.93 & 68.99 \\ \hline
0.9 & -0.5994  & -0.09178 & -1.474           & -0.3043 & -11.14 & 28.91 \\ \hline
1.0 & -0.02956 & -1.843   &  0.306          & -0.1070 & 1.25 & -3.17 \\ \hline
1.1 & 1.475    & 0.3780   & 1.360        & 0.7565 & 13.26 &-33.10 \\ \hline
1.2 & 6.375    & 0.5344   & 0.442     & 3.214 & 22.90 & -56.33 \\ \hline
1.3 & -3.090   & 0.4390   & -0.923        & -1.523 & -15.32 & 40.04 \\ \hline
1.4 & -0.2568  & -0.1551  & -0.858       & -0.1362 & -6.23 & 16.04 \\ \hline
1.5 & 0.003605 & -0.6429  &  0.351         & -0.03034 & 2.01 & -4.45 \\ \hline
1.6 & 1.281    & 0.2144   &  0.935         & 0.6511 & 9.93 & -24.93 \\ \hline
1.7 & 4.958    & 0.2672   &  0.186               & 2.493 & 16.78 & -41.67 \\ \hline
1.8 & -1.903   & 0.2196   & -0.732         & -0.9030 & -10.24 & 26.54 \\ \hline
1.9 & -0.1193  & -0.1560  & -0.557      & -0.06743 & -3.91 & 10.03 \\ \hline
2.0 & 0.02723  & -0.2726  & 0.360        & -0.000014 & 2.25 & -5.73
\end{tabular}
\end{table}

\begin{table}
\noindent Table 2. Dependence of the transition rate and the life-time on
 $Z$ for $b=1.2\mu m$. $\w0 =1.5\times 10^{16}$ Hz.
\begin{tabular}{ccccc}
$Z/b$ & $\Delta_0 \times 10^2 $ & $\Delta (b) \times 10^2$ & $\Delta W$(MHz) & $\Delta\tau$(ps) \\ \hline
0    & 0.442  & 3.214   & 22.90  & -56.33 \\ \hline
1/20 & 0.841  & 2.811   & 22.87  & -56.27 \\ \hline
1/19 & 0.459  & 3.202   & 22.93  & -56.40 \\ \hline
1/18 & -0.0233& 3.441   & 21.40  & -52.78 \\ \hline
1/17 & -0.564 & 3.435   & 17.98  & -44.57 \\ \hline
1/16 & -1.075 & 3.088   & 12.61  & -31.51 \\ \hline
1/15 & -1.410 & 2.349   & 5.88   & -14.86 \\ \hline
1/14 & -1.374 & 1.289   & -0.532 &  1.36  \\ \hline
1/13 & -0.791 & 0.2083  & -3.65  &  9.36 \\ \hline
1/12 & 0.306  & -0.2960 & 0.06   & -0.16 \\ \hline
1/11 & 1.375  & 0.4584  & 11.48  & -28.75 \\ \hline
1/10 & 1.265  & 2.376   & 22.80  & -56.10 \\ \hline
1/9  & -0.618 & 3.347   & 17.09  & -42.42 \\ \hline
1/8  & -1.478 &0.9838   & -3.10  & 7.93 \\ \hline
1/7  &  1.528 &0.4776   & 12.56  & -31.40 \\ \hline
1/6  & -1.303 &2.974    & 10.47  & -26.25 \\ \hline
1/5  &  2.088 &1.486    & 22.38  & -55.11
\end{tabular}
\end{table}


\begin{references}
\bibitem{ref1} H.\ B.\ G.\ Casimir, Proc.\ Ned.\ Akad.\ Wet {\bf 51},
    793 (1948);
    J.\ Chem.\ Phys.\ {\bf 46}, 407 (1949).
\bibitem{ref2} Il-T.\ Cheon, Phys.\ Rev.\ {\bf {A37}}, 2785 (1988).
\bibitem{ref3} G.\ Barton, Proc.\ Roy.\ Soc.\ London {\bf {A410}},
    175 (1987).
\bibitem{ref4} G.\ Plunien, B.\ M\"{u}ller and W.\ Greiner, Phys. Reports
{\bf 134}, 85 (1986).
\bibitem{ref5} A.\ I.\ Ferguson and J.\ M.\ Tolchard, Contemporary Phys. {\bf 28},
383 (1987).
\bibitem{ref6} D.\ Meschede, Phys.\ Reports {\bf 211}, 201 (1992). Other related
papers can be found in references of this paper.
\bibitem{ref7} J.\ J.\ Sakurai, ``Advanced Quantum Mechanics", Addison-Wesley
    Pub.\ Co.\ 1976. Chapt. 2,
    L.\ I.\ Schiff, ``Quantum Mechanics", 3rd.\ ed.\ McGraw-Hill, 1968.
    Chapt. 10.
\bibitem{ref78} W.\ H.\ Louisell, "Quantum Statistical Properties of Radiation",
John wiley \& Sons, New York, 1973. Chap. 1.
\bibitem{ref8} E.\ A.\ Hildum, U.\ Boesl, D.\ H.\ McIntyre, R.\ G.\ Beausoleil
    and T.\ W.\ H\"ansch, Phys.\ Rev.\ Lett.\ {\bf {56}}, 576 (1986).
\bibitem{ref9} Yu.\ L.\ Sokolov and V.\ P.\ Yakovlev, Zh.\ Eksp.\ Teor.\ Fiz.\
    {\bf {83}}, 15 (1982)
    [Sov.\ Phys. - JETP {\bf {56}}, 7 (1982)].
\bibitem{ref15} K.\ Kakazu and Y.\ S.\ Kim, Phys.\ Rev.\ {\bf A50}, 1830 (1994).
\bibitem{ref10} V.\ M.\ Mostepanenko and N.\ N.\ Trunov, Soviet Physics
Uspekii {\bf 31}, 965 (1988).
\bibitem{ref11} M.\ Kreuzer and K.\ Svozil, Phys.\ Rev.\ {\bf D34},
1429 (1986).
\bibitem{ref12} Il-T.\ Cheon, J.\ Phys.\ Soc.\ Japan {\bf 61}, 1535 (1992).
\bibitem{ref13} M.\ Bordag, D.\ Robaschik and E.\ Wieczorek, Ann.\ Phys.\
{\bf 165}, 192 (1985).
\bibitem{ref16} M.\ R.\ Philpott, Chem.\ Phys.\ Lett.\ {\bf 19}, 435(1973); J.\
Chem.\ Phys. {\bf 62}, 1812(1975).
\bibitem{ref17} G.\ Barton, Physica Scripta {\bf T21}, 11(1988).
\bibitem{ref18} Il-T.\ Cheon, J.\ Korean Phys.\ Soc. {\bf 29}, 247(1996).
\end{references}
\end{document}